\documentclass{aa}  

\usepackage{graphicx}
\usepackage{txfonts}
\usepackage{epstopdf}
\usepackage{longtable}
\begin{document}

   \title{The Solar Twin Planet Search}

   \subtitle{III. The [Y/Mg] clock: estimating stellar ages of solar-type stars\thanks{Based on observations obtained at the
    Clay Magellan Telescopes at Las Campanas Observatory, Chile and at the 3.6m Telescope at the La Silla ESO Observatory, Chile (program ID 188.C-0265).}
    }

   \author{M. Tucci Maia
          \inst{1}
          \and
          I. Ram{\'{\i}}rez\inst{2}
          \and
          J. Mel\'{e}ndez\inst{1}
          \and 
          M. Bedell\inst{3}
          \and
          J. L. Bean\inst{3}
          \and
          M. Asplund\inst{4}
          }

   \institute{Universidade de S\~ao Paulo, Departamento de Astronomia do IAG/USP, Rua do Mat\~ao 1226, 
              Cidade Universit\'aria, 05508-900 S\~ao Paulo, SP, Brazil. \email{marcelotuccimaia@usp.br}
         \and
             University of Texas, McDonald Observatory and Department of Astronomy at Austin, USA
         \and
             University of Chicago, Department of Astronomy and Astrophysics, USA
         \and
             The Australian National University, Research School of Astronomy and Astrophysics, Cotter Road, Weston, ACT 2611, Australia}

   \date{Received ... 2015; accepted ... }

 
  \abstract
   {Solar twins are stars with similar stellar (surface) parameters to the Sun that can have a wide range of ages. This provide an opportunity to analyze the variation of
   their chemical abundances with age. Nissen (2015) recently suggested  that the abundances of the s-process element Y and the $\alpha$-element Mg could be used to estimate stellar ages.}
   {This paper aims to determine with high precision the Y, Mg, and Fe abundances for a sample of 88 solar twins that span a broad age range ($0.3-10.0$\,Gyr) and investigate their use
   for estimating ages.}
   {We obtained high-quality Magellan Inamori Kyocera Echelle (MIKE) spectra and determined Y and Mg abundances using equivalent widths and a line-by-line differential method within a 1D LTE framework. 
   Stellar parameters and iron abundances were measured in Paper I of this series for all stars, but a few (three) required a small revision.}
   {The [Y/Mg] ratio shows a strong correlation with age. It has a slope of -0.041$\pm$0.001 dex/Gyr and a significance of 41 $\sigma$.
   This is in excellent agreement with the relation first proposed by \cite{nis15}. We  found some outliers that turned out to be binaries
   where mass transfer may have enhanced the yttrium abundance. Given a precise measurement of [Y/Mg] with typical error of 0.02 dex in solar twins, our 
   formula can be used to determine a stellar age with $\sim$0.8 Gyr precision in the 0 to 10 Gyr range.}
   {}


   \maketitle
%

\section{Introduction}

Solar twins are stars that have spectra very similar to the Sun, with stellar (surface) parameters (temperature, surface gravity, and metallicity) around the solar values (T$\rm_{eff}$ within $\pm 100$ K, log $g$ and 
[Fe/H] within $\pm$ 0.1 dex, as arbitrarily defined in \citep{ram14})\footnote{We note that some stars in \cite{ram14} fall slightly outside the solar twin definition. They are also included in this work because they are close
enough to the Sun for a high-precision abundance analysis.}. Since they have about 1 M$_{\rm \odot}$ 
and roughly solar chemical composition,
they follow a similar evolutionary path to the Sun, from the zero age main sequence to the end of their lives. The highly precise atmospheric parameters that one can derive for these objects allows 
a reliable determination of their ages using the traditional isochrone method \citep{ram14,nis15}. Thus, we can take advantage of this very special group of stars to better understand the nucleosynthesis of s- and 
r-elements throughout the Galaxy \citep[e.g.][]{mas00,bat15}. 
        
Another important potential application of the heavy elements is their use for age dating. By investigating the abundances of several elements using high precision differential
abundances for a sample of 21 solar twins, \cite{nis15} finds a very tight correlation of [Y/Mg] as a function of stellar age. There have also been previous studies at standard precision that indicate a 
correlation between the s-process elements, like Ba and Y, with stellar age \citep{mas00,ben05,dor09}. More recently, \cite{mai11} reinforced the above results using open clusters in a broad age range.
        
The aim of this work is to analyze the abundances of the heavy element yttrium and the $\alpha$-element magnesium in a sample of 88 solar twins with ages covering
0.3 Gyr to 10.0 Gyr, which thus have important implications for astronomy, such as for dating exoplanet host stars, studying stellar evolution 
effects, Galactic chemical evolution, and different studies of stellar populations.



\section{Data and analysis}
\subsection{Observations and data reduction}

The observations for the 88 stars of our sample of solar twins were carried out with the Magellan Inamori Kyocera Echelle (MIKE) spectrograph \citep{ber03} on the 6.5m Clay Magellan Telescope at Las Campanas Observatory on five runs between
January 2011 and May 2012. See \cite{ram14} for a more detailed description of our sample, the observations, and data reduction.

The same instrumental setup was employed for all stars, achieving a S/N ratio of at least 400 around 600 nm. The resolving power is R = 83000 in the blue, and
R = 65000 in the red. The spectra of the Sun, which served as a reference for the differential analysis, were obtained through the observation of the asteroids Iris and Vesta using the same instrumentation
setup\footnote{In this work we only use the light reflected on Iris as our reference spectrum for the differential analysis.}. 
The orders were extracted with the CarnegiePython MIKE pipeline\footnote{http://code.obs.carnegiescience.edu/mike},
and Doppler correction and continuum normalization was performed with IRAF.

\subsection{Stellar parameters}

Stellar parameters were obtained by \cite{ram14} through differential excitation and ionization equilibrium using the abundances of FeI and FeII, with the Sun as reference.  
The abundances were determined using the line-by-line differential method, employing EW that were measured by hand with the task splot in IRAF. The Fe abundances and stellar parameters from \cite{ram14},
were determined with the 2014 version of the LTE code MOOG \citep{sne73}, adopting the MARCS grid of 1D-LTE model atmospheres \citep{gus08}.
The Y and Mg abundances were determined with same grid, but we note that the exact grid of model atmospheres is irrelevant for differential abundances, since the mean abundance difference is lower than 0.001 dex \citep{mel12}.
We have repeated our calculations using  Kurucz ODFNEW model atmospheres \citep{cas04} and found that the mean differential [Mg/H] and [Y/H] abundance change in only 
(Kurucz - MARCS) -0.0006 dex ($\sigma = 0.0031$) and -0.0003 dex ($\sigma = 0.0045)$, respectively.


We also employed the recently introduced python q${\rm^2}$ code\footnote{https://github.com/astroChasqui/q2; a tutorial is available on this site with 
detailed information on the capabilities of this code.} \citep{ram14}, which makes the abundance determination and analysis considerably more efficient, 
by calling MOOG drivers and performing the line-by-line analysis, including corrections by hyperfine structure (HFS) and also
computing the associated errors. Both observational and systematic uncertainties were considered. Observational errors are due to uncertainties in the measurements (standard error) while the systematic 
errors are uncertainties coming from the stellar parameters, as described in \cite{ram15}. Observational and systematic errors were added in quadrature.

The age and mass for the sample were determined using Yonsei-Yale isochrones \citep{yi01}, as described in \cite{ram13,ram14}. 
This method provides good relative ages, due to the high precision of the atmospheric
parameters, by comparing the location of the star on the T$_{\rm eff}$, log $g$, [Fe/H] parameter space, with the values
predicted by the isochrones, computing mass, and age probability distribution functions. As shown below, these ages can also be made accurate (i.e. 
almost insensitive to the choice of models) by forcing different isochrone sets to reproduce  the solar parameters exactly.

\begin{figure}
\centering
\includegraphics[width=1.0\columnwidth]{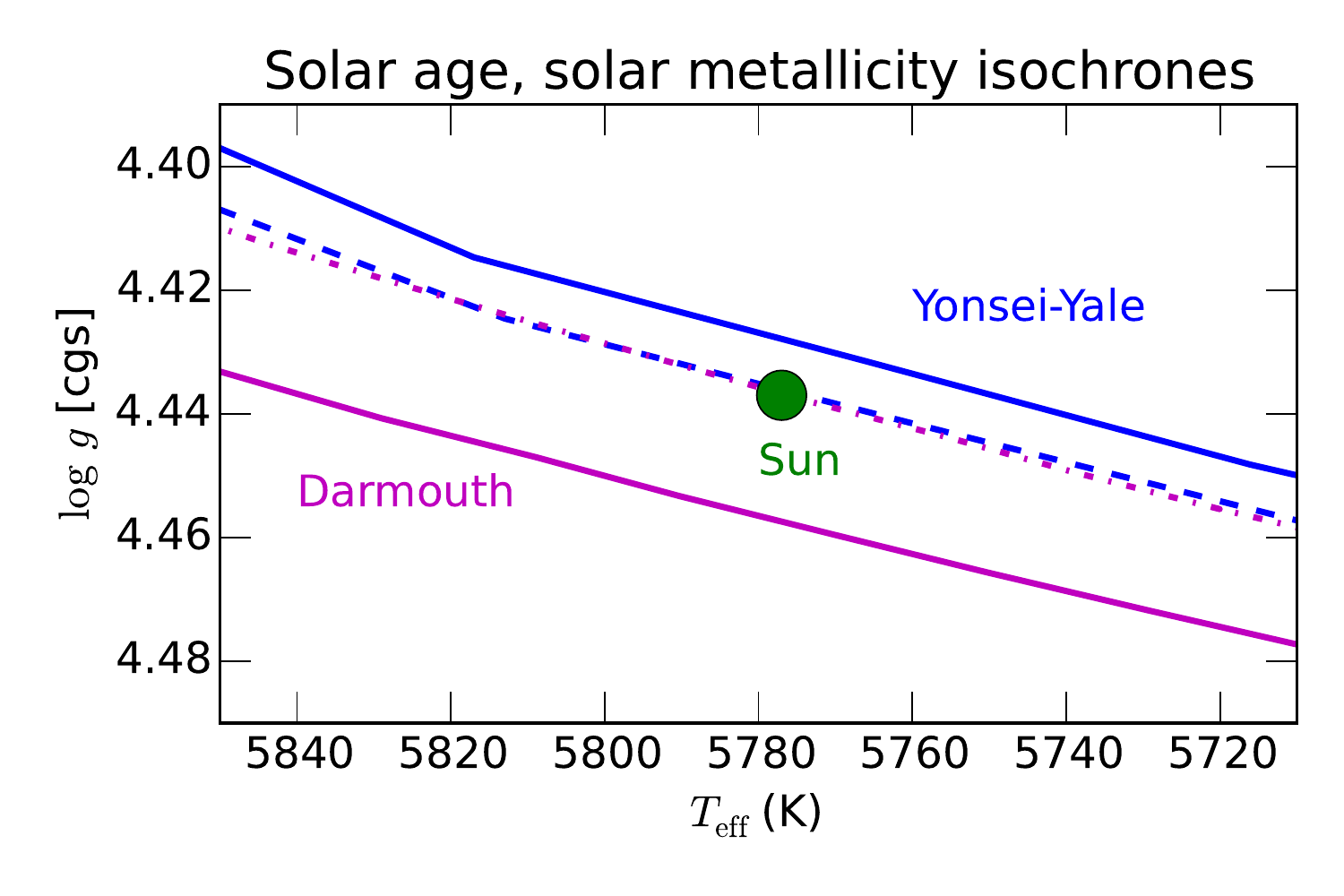}
\caption{The 4.6 Gyr Yonsei-Yale (dashed line) and 4.5 Gyr Darmouth (dot-dashed line) isochrones shifted in [Fe/H] by -0.04 dex and +0.08 dex, respectively. We note the 
agreement after the change.}
\label{iso_sun}
\end{figure}

Figure \ref{iso_sun} shows the location of the Sun in the T$\rm_{eff}$-log $g$ plane along with 4.6 Gyr Yonsei-Yale (YY) and 4.5 Gyr Darmouth (DM) isochrones \citep{dot08}.
These ages are the closest to solar age found in each grid. Solid lines represent the isochrones of solar composition in each case.
Clearly, they do not exactly pass through the solar location, but a minor shift of the [Fe/H] of the isochrone sets by -0.04 in the case of
YY (dashed line) and +0.08 for DM (dot-dashed line) brings these isochrones to excellent agreement with the solar parameters at the well-known solar age \citep[e.g.][]{sac93}. 
We applied these offsets to both isochrone grids before using them to determine stellar parameters.\footnote{in R14, the -0.04 dex offset in the 
YY isochrone [Fe/H] values was applied after selecting the isochrone points to use in the probability density (PD) calculations. 
This led to a very minor offset ($-0.1\pm0.2$ Gyr) in the ages derived with respect to the more precise case where the isochrone [Fe/H] 
values are all shifted before selecting the points to use in the PD computation. This minor change makes the ages reported in R14 slightly 
different from those employed in this work, but these small differences do not affect the results presented in this paper in any significant 
way.}

\begin{figure}
\centering
\includegraphics[width=1.0\columnwidth]{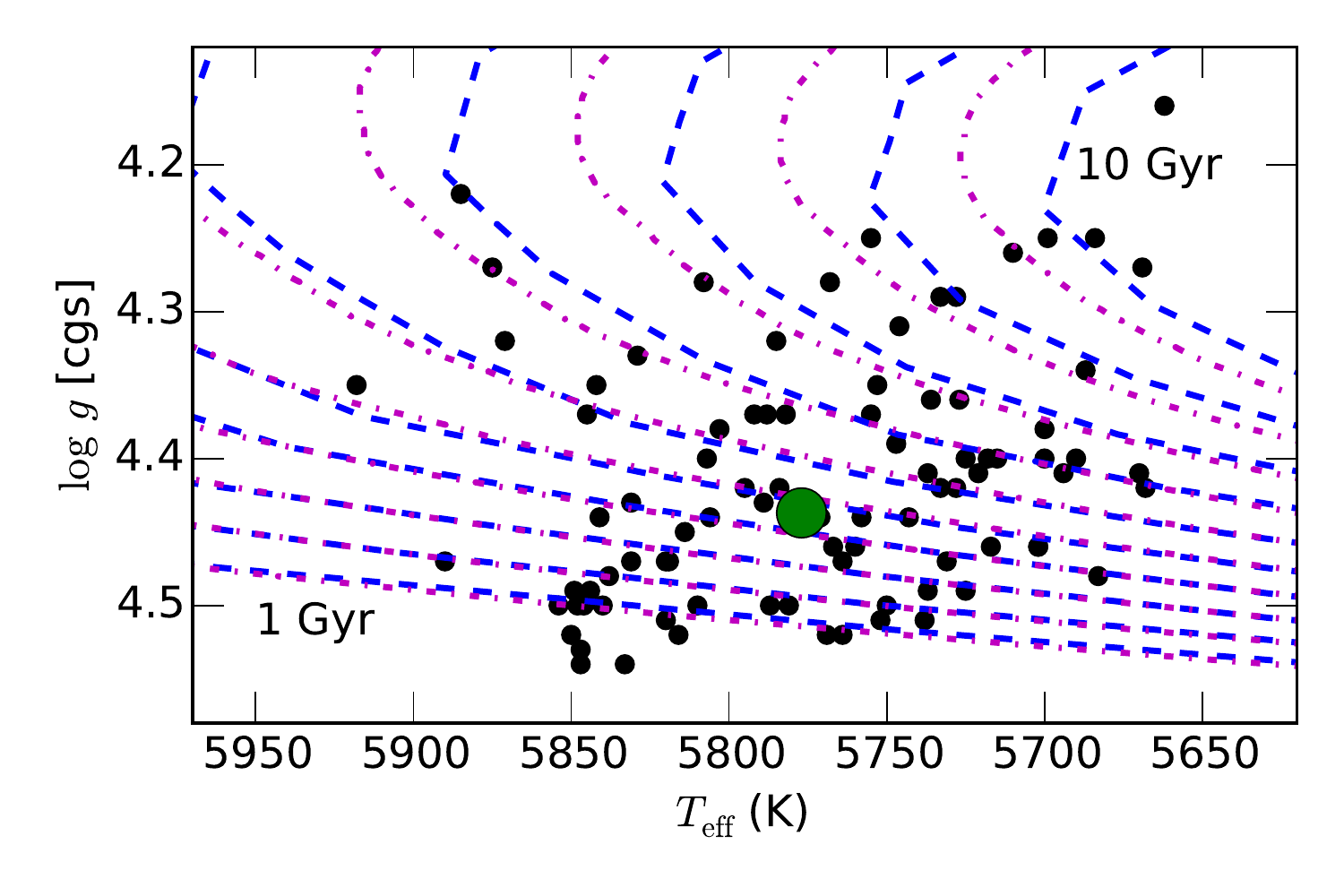}
\caption{Distribution of our sample with the 1 to 10 Gyr Yonsei-Yale
(dashed lines) and Darmouth (dot-dashed lines) isochrones.}
\label{iso_stps}
\end{figure}

As shown in Figure \ref{iso_stps}, our solar twin data set spans a narrow range of T$_{\rm eff}$ and log$g$, but it is enough to cover 
the very wide range of ages from 0 to ~10 Gyr. YY (dashed lines) and DM (dot-dashed lines) isochrones are also 
shown in this plot. These isochrones have [Fe/H]= -0.04 for Yonsei-Yale and [Fe/H]=+0.08 for DM, which, as explained above, 
pass through the solar location at solar age. We note the excellent agreement between these two sets of isochrones for ages younger than 6 Gyr.
For older stars, the DM isochrones are shifted to somewhat higher effective temperatures, which implies that the ages inferred 
from them will be somewhat older, compared  to those obtained from the Yonsei-Yale set.

\begin{figure}
\centering
\includegraphics[width=1.0\columnwidth]{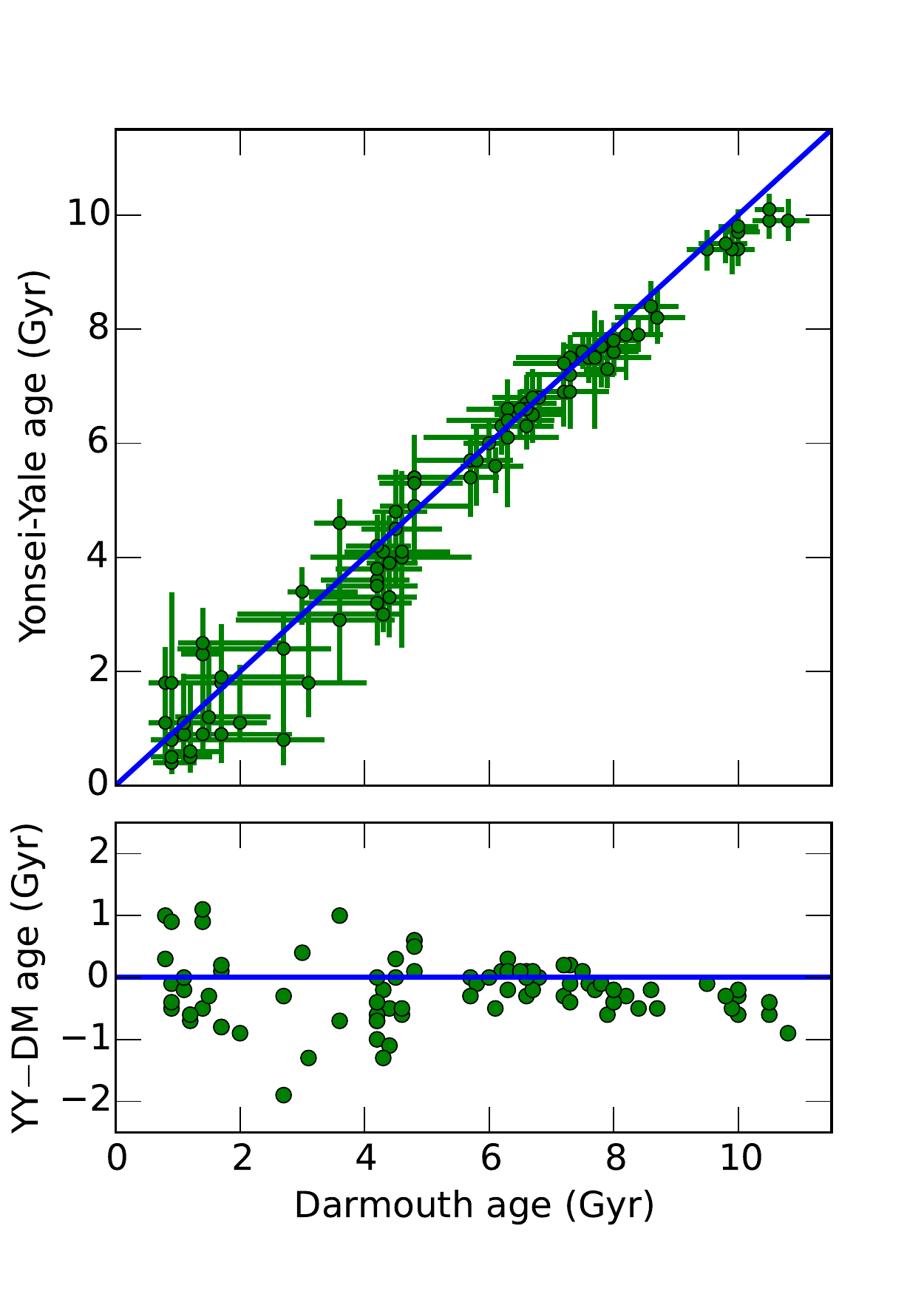}
\caption{In the upper panel, we make a comparison of ages estimated by Yonsei-Yale and Darmouth isochrones for our solar twin sample.
The lower panel shows the differences between the YY and DM isochronal ages.}
\label{agecomp}
\end{figure}

Figure \ref{agecomp} compares the YY and DM isochrone ages derived for our solar twin stars. On average, 
the mean difference of the most probable ages (DM--YY) is $+0.2\pm0.5$ Gyr, which would suggest good agreement within the errors. However, 
there is a clear systematic offset at older ages, albeit small, of $+0.4\pm0.2$ Gyr.

If the [Fe/H] offsets to the isochrones are not applied, the YY and DM isochrones are systematically off by 1 Gyr at solar age, and up to 2 Gyr for the oldest stars.
On the other hand, when these corrections are applied to the isochrones, the anchor points are the solar parameters, which give us relative accurate ages. We note that, for the pair 
of old solar twins 16 Cyg, our method gives an age of $7.1^{+0.2}_{-0.4}$ Gyr 
(from the combined age-probability distributions; \cite{ram11}), which is in excellent agreement
with the seismic ages recently determined for this pair (average of 7.0 $\pm$ 0.1 Gyr; \cite{san16}).
Even though the typical error for both isochronal ages set is $\sim$ 0.6 Gyr, we decided to use 
the YY grid instead of DM because the former has a more consistent sampling of the isochrones, which makes the age determination less likely to suffer from statistical biases. 

The stellar parameters and [Fe/H] abundances for most of our sample stars were determined in our previous work \citep{ram14},
except for HIP 108158, HIP 55409, HIP 72043, and HIP 68468. As these stars were outliers in the [Y/Mg] versus age plot 
(there are other outliers, but they can be explained owing to to binarity), we decided to verify their parameters by
remeasuring the EW of FeI and FeII lines for those stars (the reanalysis of HIP 68468 is presented in \cite{mel16}). For HIP 72043, we didn’t find any 
difference, meaning that it is a true outlier in the [Y/Mg]-age plane; for the other three stars, their parameters were revised (Table \ref{par}).


\begin{table}
\caption{Revised parameters for HIP 108158, HIP 55409, and HIP 68468.}
\small
\label{par}
{\centering
\renewcommand{\footnoterule}{}
\begin{tabular}{lccccl} 
\hline\hline                
 {Star}& T${\rm_eff}$ & log g & [Fe/H] & Mass &  Age\\
 {HIP} & (K) & (dex) & (dex) & (M$\odot$) & (Gyr)\\
\hline 
108158 & 5688$\pm$6 & 4.29$\pm$0.02& 0.067$\pm$0.008 & $0.99^{\rm 1.01}_{\rm 0.98}$ & $9.0^{\rm +0.4}_{\rm -).4}$\\
55409  & 5712$\pm$6 & 4.41$\pm$0.02& -0.060$\pm$0.006 & $0.96^{\rm 0.97}_{\rm 0.95}$ & $6.9^{\rm +0.7}_{\rm -0.7}$\\
68468  & 5857$\pm$8 & 4.32$\pm$0.02& 0.065$\pm$0.007 & $1.05^{\rm 1.06}_{\rm 1.04}$ &$5.9^{\rm +0.4}_{\rm -0.4}$\\

\hline                                 
\end{tabular}
}
\end{table}

We also updated the ages for HIP 109110 and HIP 29525, two young solar twins, for which more precise ages, which were  determined through rotational periods, are available in \cite{bau10}\footnote{We note that
HIP 109110 is not used in the linear fit because it was identified as a spectroscopic binary. For HIP 29525, even if we adopt the more uncertain isochronal age, the linear
fit is not changed because of the large error bar in age for this star.}. According
to \cite{bar07}, the errors from gyrochronology is 15\% in the age of solar analogs, which is significantly  
better than those  found in \cite{ram14},which are about 40-70\% for these two young stars (isochrone ages have larger error bars at a younger age, as seen in Figure \ref{agecomp}).
We note also that, for those two stars, the rotational ages agree better with the [Y/Mg] ages.

\subsection{Abundance analysis}

Yttrium abundances were obtained using the 485.48nm, 520.04nm, and 540.27nm YII lines and corrected for HFS adopting the HFS data from Meléndez et al. (2012).

\begin{figure}
\centering
\includegraphics[width=1.0\columnwidth]{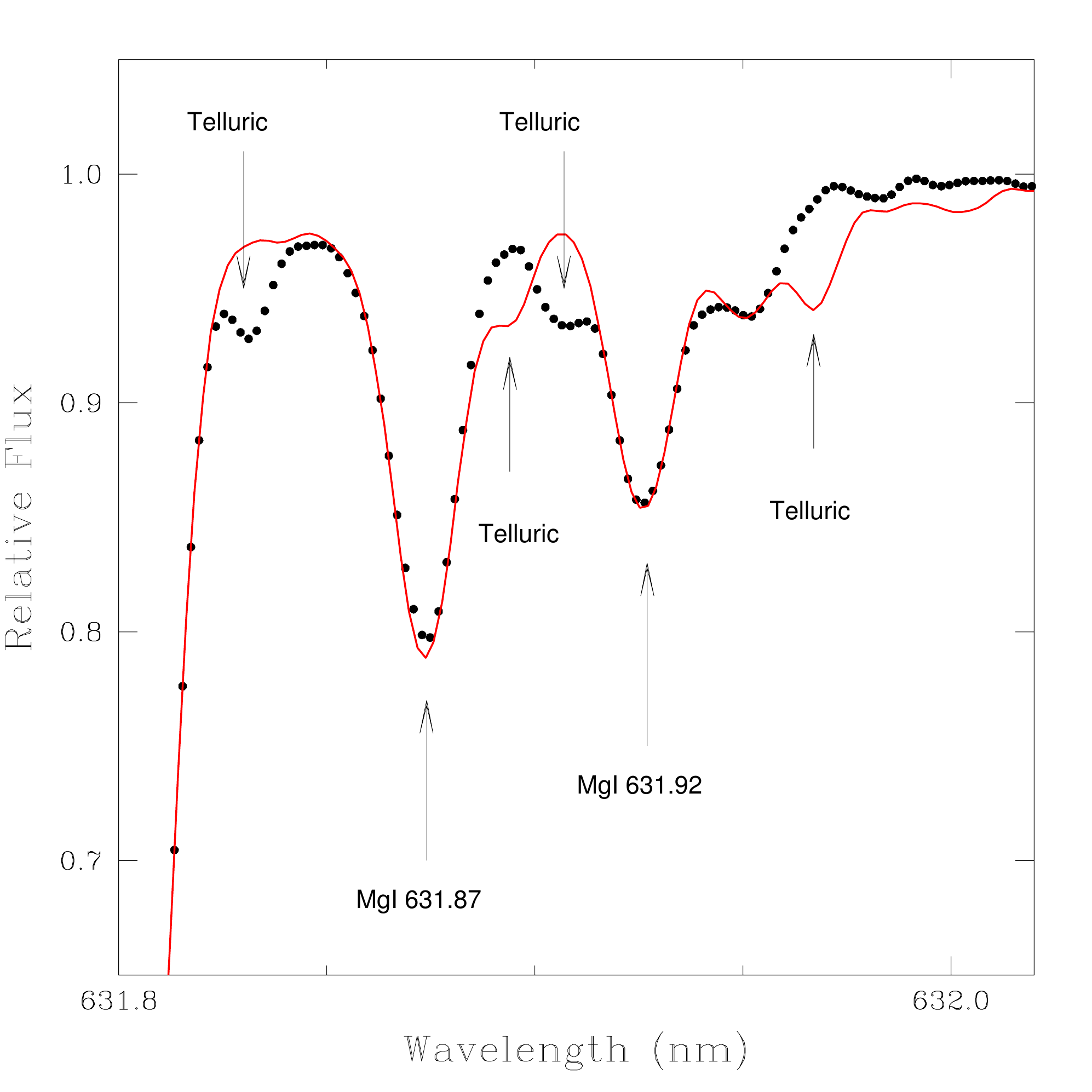}
\caption{Mg I lines around 631.9 nm and telluric lines in this region for HIP 64713 (black dots) and HIP 89650 (red line). 
We advise caution swhen measuring these Mg I lines.}
\label{espec}
\end{figure}

For magnesium we used the 454.11nm, 473.00nm, 571.11nm, 631.87nm, and 631.92nm lines, paying extra attention to the latter two lines owing to the influence of telluric features in this region, as shown in Fig. \ref{espec}.
We note that the separation between theses telluric lines is 0.74 \AA \, and their line ratio is 1.05.

Once the initial set of differential abundances was obtained, we verified the presence of outliers and, when present, the EW of those lines were verified and q$^{\rm 2}$ was executed again.     

\section{Results and discussions}

\begin{figure}
\centering
\includegraphics[width=1.0\columnwidth]{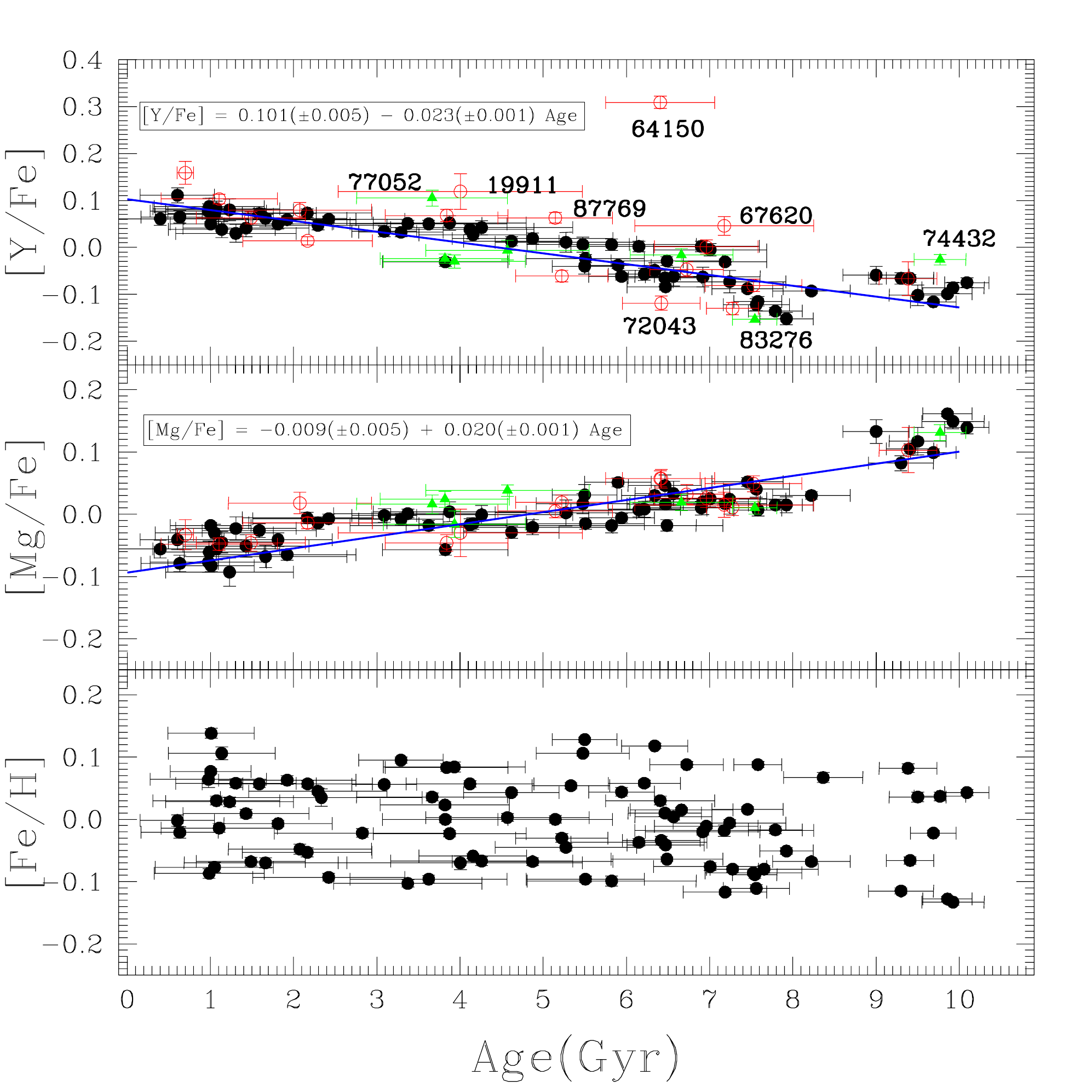}
\caption{[Y/Fe] (upper panel), [Mg/Fe] (middle panel) and [Fe/H] (lower panel) as function of age. 
The red open circles are spectroscopic binary stars and the green triangles are visual binaries. 
[Fe/H] vs. age do not show any correlation with age.
We also present the linear fit for [Y/Fe] and [Mg/Fe] versus age on their respective panels.}
\label{fig1}
\end{figure}

As shown in Fig. \ref{fig1}, there is a clear correlation between both [Y/Fe] and [Mg/Fe], and stellar age, for the sample of 88 stars, confirming the result 
found by \cite{nis15}, which is based on a smaller sample. The behaviour of yttrium is due to the increasing contribution of s-process elements from low and intermediate mass AGB stars, which most efficiently produce Y \citep{fis14,ka14} and 
which slowly became more important with time \citep{tra04,nis15}.

On the other hand, the correlation of [Mg/Fe] with age is an effect of the increasing number of Type Ia SNe in comparison to the
number of Type II SNe, as discussed by \cite{kob06}. This is because Type II SNe produces mainly $\alpha$-elements (O, Mg, Si, S, Ca, and Ti), enhancing the interstellar medium
with these species in the early Galaxy, while Type Ia SNe produce yields with high Fe/$\alpha$ ratio. Complementary to this, we show with the [Fe/H] vs. age
plot that there is no age-metallicity correlation for the stars in our data, independently of its population (Fig. \ref{fig1}).

In Fig. \ref{fig1}, there is a gap around 8.5 Gyrs that could be important when distinguishing different populations. This gap in the [Mg/Fe] vs. age plot 
was used to identify 10 stars that display a high-$\alpha$ abundance that, according to \cite{hay13}, may belong to the thick disk population.
On the other hand, \cite{adi11} classify these high-$\alpha$ metal-rich stars (h$\alpha$mr) as being a different population of stars, that do not belong to either the thin or the thick disk.
These stars share some properties with both thin and thick disks stars and might have migrated from the inner parts of the Galaxy \citep{adi13}. However the detailed study of these stars by \cite{hay13} indicates
that they may have formed at the end of the thick disk.

\begin{figure}
\centering
\includegraphics[width=1.0\columnwidth]{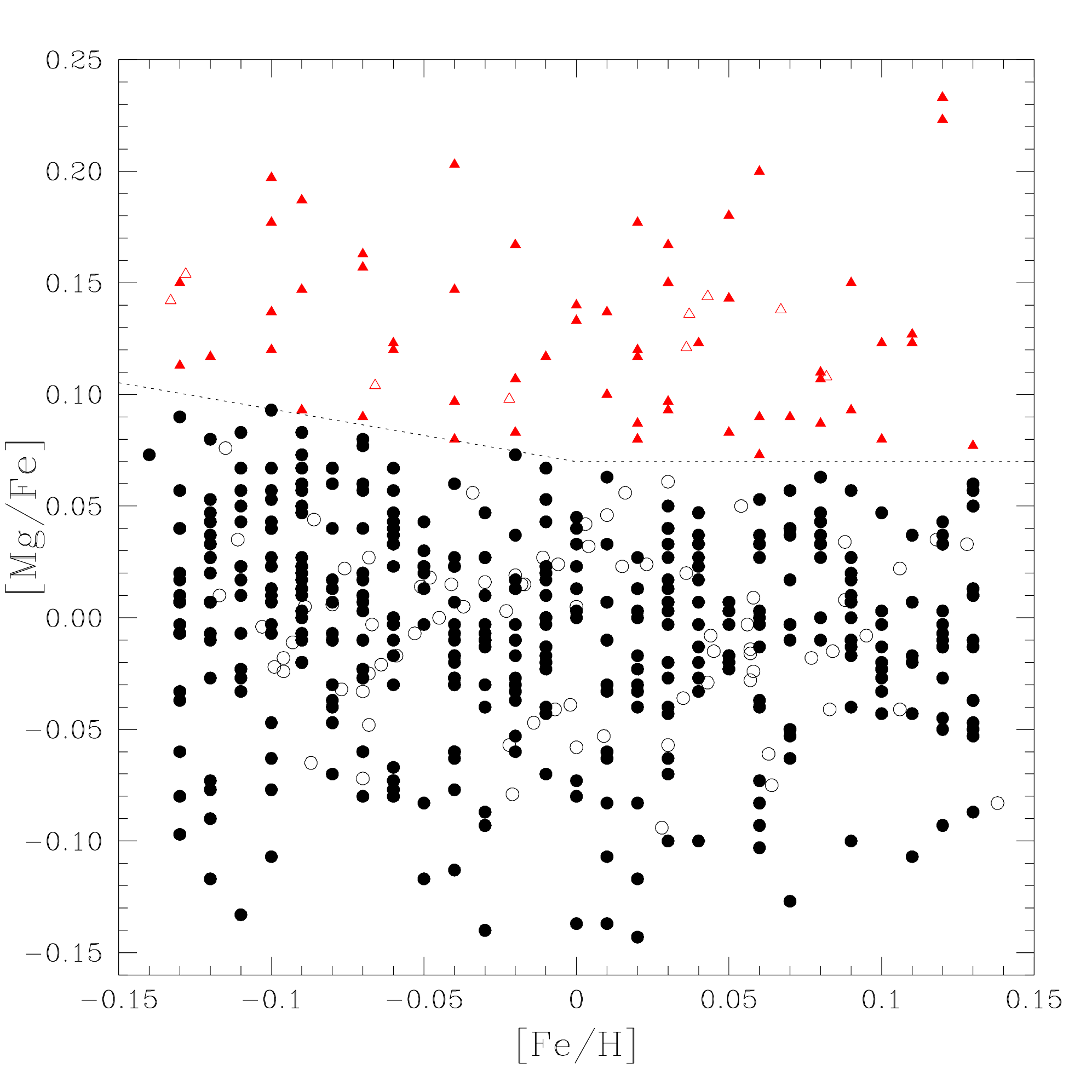}
\caption{[Mg/Fe] vs. [Fe/H] plot of the data from \cite{adi12} (filled symbols) matching  the [Fe/H] range from our work (empty points). The circles represents the thin disk stars and the 
triangles the h$\alpha$mr stars.}
\label{alpha}
\end{figure}

Using the \cite{adi13} criteria, nine stars from our sample are h$\alpha$mr (Fig. \ref{alpha})\footnote{Using data from \cite{adi12}, that matches the range of metallicities of our
sample (-0.140 to 0.140 dex). Notice that we use [Mg/Fe] rather than [$\alpha$/Fe] (which is the average of Mg, Si and Ti).}. With this method, we identify the same h$\alpha$mr stars as we did using the [Mg/Fe] vs. age plot (Fig. \ref{fig1}), 
with the exception of HIP 109821.  Adopting a binomial distribution \citep[e.g.][Chapter 3]{bev69}. 
the occurrence of h$\alpha$mr in our sample is 10/88 (11.4$\pm$ 3.4\%) which is consistent with the 3/21 (14.3$\pm$7.6\%)
from \cite{nis15} and 60/413 (14.5$\pm$1.7\%) from \cite{adi12}, using the same metallicity range of our sample.

\begin{figure}
\centering
\includegraphics[width=1.0\columnwidth]{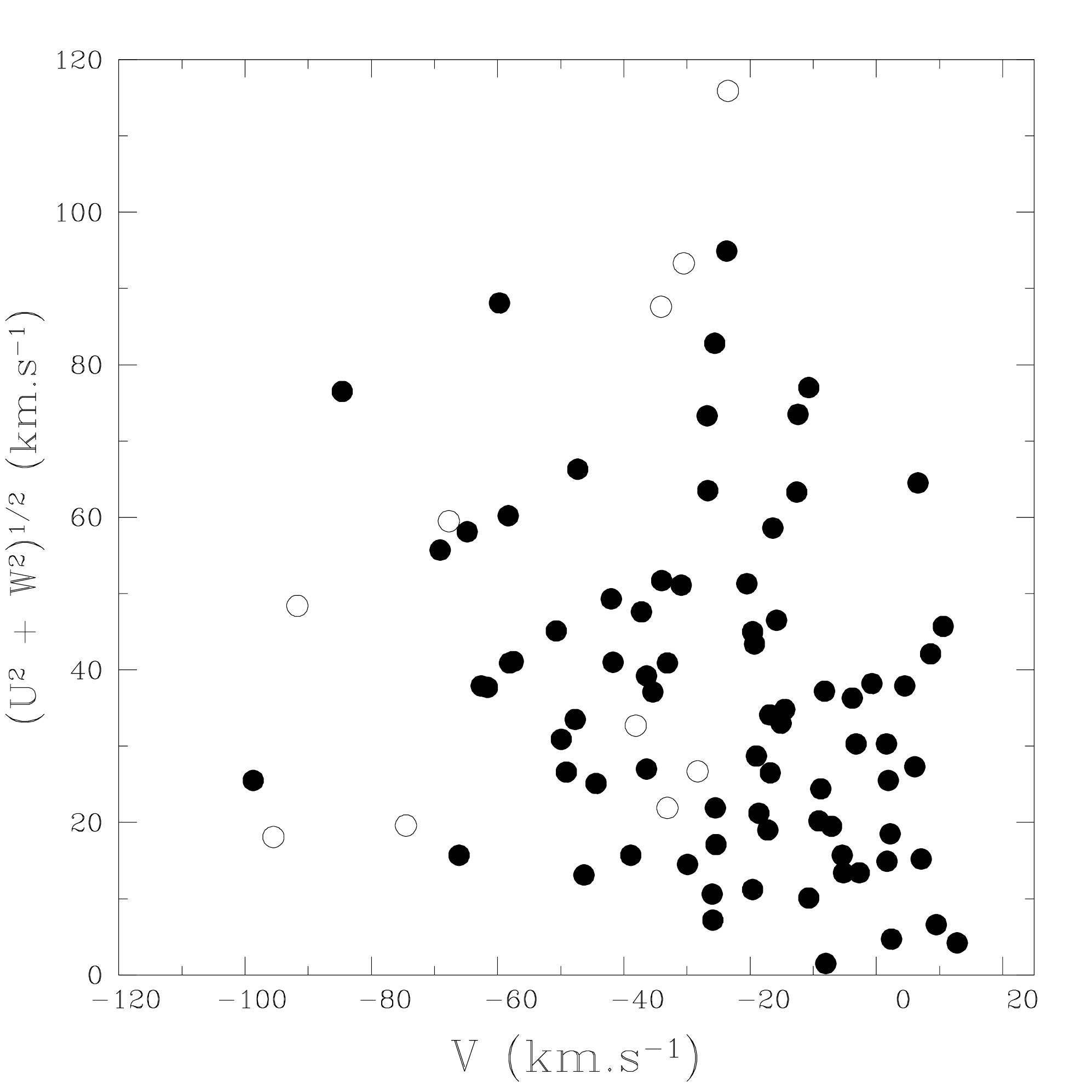}
\caption{Toomre diagram for our sample. The open circles are the high-$\alpha$ metal rich star stars.}
\label{toomre}
\end{figure}

We also show the Toomre diagram for the sample (Fig. \ref{toomre}) highlighting the h$\alpha$mr stars (open circles). The h$\alpha$mr group  does not seem to be particularly separated
from the rest of the group and its kinematic properties are in agreement with the findings by  \cite{adi11,adi13}, as well as \cite{ben14}.

The red open circles on the [Y/Fe] and [Mg/Fe] plots (and in Fig. \ref{fig3}) are binaries, pinpointed through radial velocity changes. The majority of the stars from our MIKE sample overlap with our HARPS Large Program \citep{ram14}, in which we search 
for exoplanets in a sample of about 60 solar twins using the HARPS spectrograph \citep{may03}. Thanks to the radial velocity data of our sample (and previous works), we identified some binary or multiple system stars (red open circles), 
indicated in our online Table 2. From our visual inspection, all spectroscopic binaries seem single-lined. Also, the single-lined nature of the spectra is apparent in the iron abundance analysis; 
the EWs do not appear to be contaminated in any significant manner.

We also note that in the [Y/Fe] plot, all outliers are spectroscopic binaries (red open dots). This is probably because their companion transfered Y material to what is now the primary star. 
Thus, [Y/Fe] seems to be a good method to identify potential multiple star systems where mass transfer has taken place, but this is possible only when precise ages are available. 
We note that the stars HIP 77052, HIP 74432, and HIP 83276, which  seem to be outliers in the [Y/Fe] plot, are identified as visual binaries \citep{tok14}. 

\begin{figure}
\centering
\includegraphics[width=1.0\columnwidth]{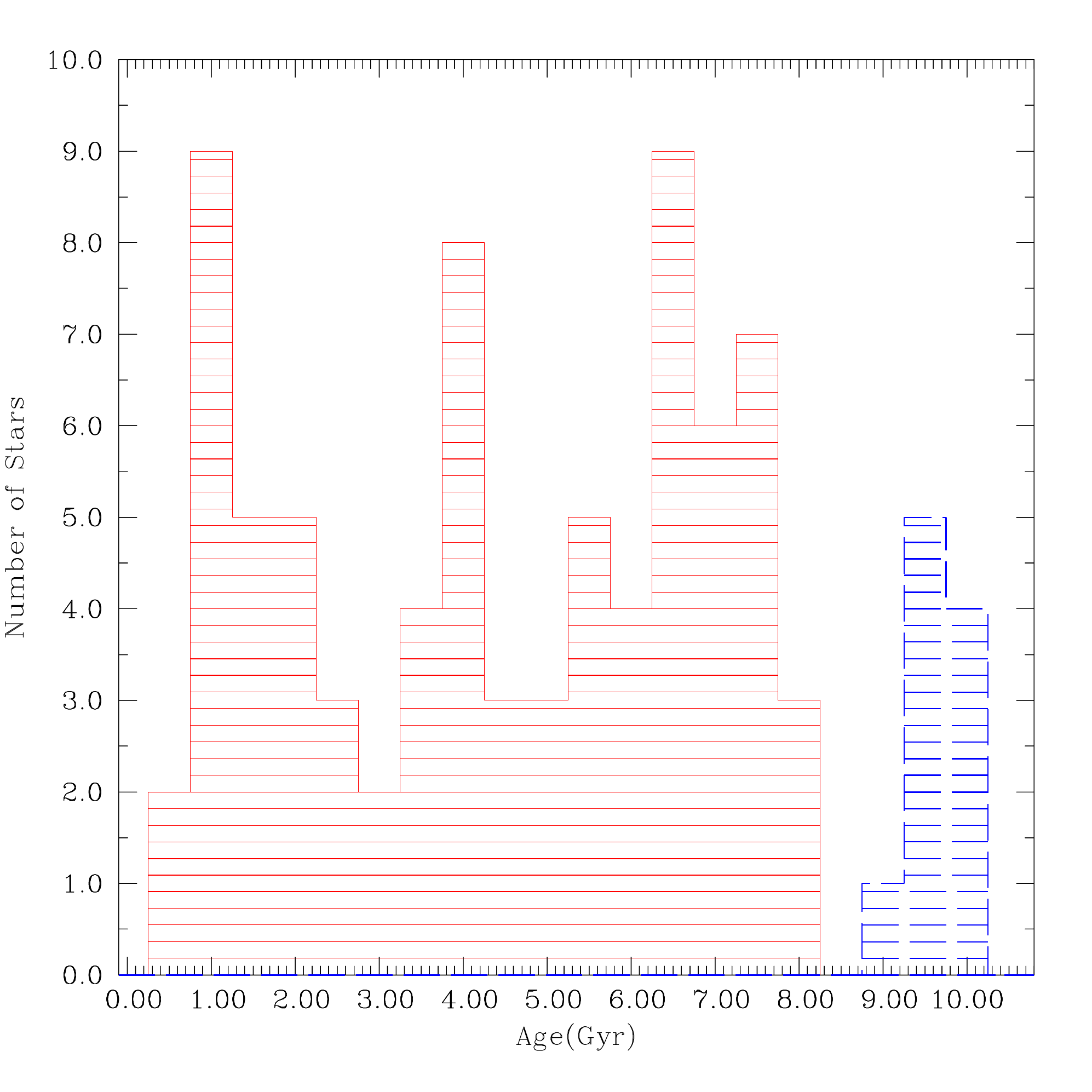}
\caption{Age histogram for the thin disk (red solid line) and h$\alpha$mr stars (blue dashed line).}
\label{fig2}
\end{figure}

\begin{figure}
\centering
\includegraphics[width=1.0\columnwidth]{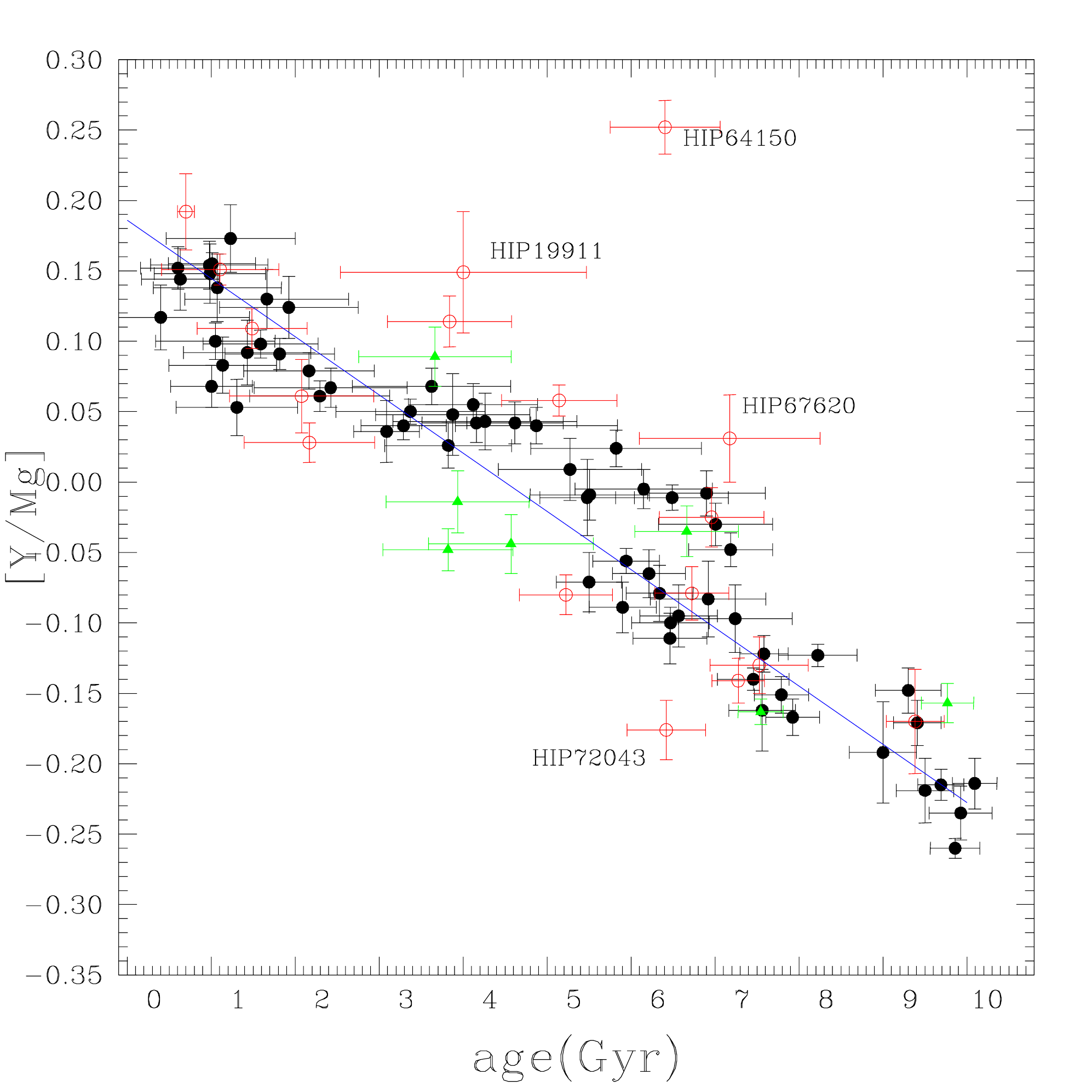}
\caption{[Y/Mg] versus age for the sample for 88 solar twins. The slope is -4.13x10$^{\rm -2} \pm 1.12$x$10^{\rm -3}$ with a scatter in age of 0.9 Gyr. The red open circles 
are spectroscopic binary stars and the green triangles are visual binaries.}
\label{fig3}
\end{figure}

Figure
 \ref{fig2} shows an age histogram of the whole sample. The red solid curve shows the thin disk, while the blue dashed line represents the 
stars assigned to the h$\alpha$mr population. It is possible to distinguish a clear age gap at 8.0 Gyrs, separating the thin disk and the h$\alpha$mr stars.
The h$\alpha$mr population show a star-to-star scatter in age of only 0.3 Gyr, showing that this population formed quickly. Nevertheless, the h$\alpha$mr stars cover an [Fe/H] 
range similar to that of the younger thin disk stars that formed during the last 8 Gyr.

In Fig. \ref{fig3} we present the [Y/Mg] vs. stellar age plot.  A linear fit to our data, excluding
the spectroscopic and visual binary stars, gives the following relation using the YY ages\footnote{We have made tests using DM and YY ages with [Y/Mg] abundances to identify which would have the better fit with age, but we found the same 
scatter (0.038 dex in [Y/Mg]) and the same slope within 1 $\sigma$. Since the differences in the age determination 
are small, they do not affect the final result.}:

\begin{equation}
$[Y/Mg]$ = 0.186(\pm 0.008) - 0.041 (\pm 0.001). $Age$ 
\end{equation}

This is practically the same fit found by Nissen (2015), within 1$\sigma$, but with better precision and a scatter of 0.037 dex. 
A remarkable significance level of about 41$\sigma$ is found for the slope and a Spearman coefficient of $r_S$ = -0.96, with a 
probability of $10\rm^{-35}$ of our results arising by pure chance,
showing that this behaviour cannot occur randomly. We note that the h$\alpha$mr stars were not excluded from the fit, 
meaning that the [Y/Mg] relation seems to also be  valid for this population.
The relation of age (in Gyr) as function of [Y/Mg] is

\begin{equation}
$Age$ = 4.50 (\pm 0.09) -24.18(\pm 0.66). $[Y/Mg]$ 
\end{equation}

The scatter of this relation is 0.9 Gyr, which is larger then the average error of the isochronal ages (0.6 Gyr). Subtracting these errors, we find an intrinsic uncertainty of 0.7 Gyr for the age 
determination (this value should be added to the error in the age determination from Eq. 1.
For [Y/Mg], we have a mean error of 0.017 dex, which translates to a typical error
in age of 0.4 Gyr. Thus, the total error expected is $\sim$ 0.8 Gyr for data with quality similar to the those employed in this work.


The age - [Y/Mg] relation shown in Fig. \ref{fig3} and described by Eqs. 1 and 2 was determined using a sample of solar twins for which highly precise Y and Mg abundances, 
as well as ages, could be derived. Its origin is explained by nucleosynthesis and the chemical evolution of the solar neighbourhood. Thus, we do not expect this 
relation to be restricted to solar twins. Since nucleosynthetic yields can be metallicity-dependent, it is possible that this relation could be different for samples with non-solar [Fe/H].
We tested this possibility by dividing our sample into metal-rich and metal-poor groups, finding no significant differences. Thus, it is possible that the metallicity 
dependency is mild, if it is present at all. This means that, at least for
the metallicity interval of our sample, the [Fe/H] does not have an impact on the age determination using Eq. 2. However, more studies are needed to determine
if this remains true to metallicities other than solar. The [Y/Mg] abundance ratio can be measured with a precision of about 0.05 dex in non-solar twins. Our Eq. 2 can be used on those
stars to determine their ages with a precision of  ~1.4 Gyr. With less precise [Y/Mg] measurements, for example assuming errors of 0.1 dex, we can still constrain the stellar age to ~2.5 Gyr.

\section{Conclusions}


We confirm the tight relation of [Y/Mg] vs. age, first found by \cite{nis15}. This relation seems to apply even for the thick disk population. Although 
we used a bigger sample of solar twins, the relation found is practically the same as Nissen's, with a slope 
of -0.041 $\pm$ 0.001 dex/Gyr and a scatter in ages of $\sigma$ = 0.9 Gyr.

The mean uncertainty expected for data with precision similar to ours is $\sim$ 0.8 Gyr. 
This level of precision for the abundances, stellar parameters, and age determination as well, could only be achieved
through a strict differential analysis of solar twin stars. We note that our careful work allowed us
to find a good correlation of [Y/Mg] abundances with stellar age. However, extremely high-precision abundances are 
not necessary to have a satisfactory age determination. Even with a [Y/Mg] ratio with error of 0.05 dex, it is
possible to obtain an age with a uncertainty of 1.4 Gyr.

Tests were made to verify if this relation has some dependence with metallicity. For this we divided the group into metal-poor and metal-rich stars, but no significant trend with
[Fe/H] was detected, meaning that the age determination, at least in the -0.14 $\leq$ [Fe/H] $\leq$ 0.14 dex interval, should not be metallicity-correlated. 

More investigation is needed to verify the applicability of the [Y/Mg] clock relation to 
stars with metallicities that are different from solar. Also, this correlation may be
more complex than just a simple linear fit.
Nevertheless, with regard to solar twins and solar analogs, the 
[Y/Mg] ratio is a promising new metric to
reliably estimate relative ages independent of isochrones, and could be used alongside  other age determination methods.

Our work provides important observational constraints to the yields of s-process elements in models of low- 
and intermediate-mass AGB stars \cite[e.g.][]{mai12}. 


\begin{acknowledgements}
MTM thanks support by CNPq (142437/2014-0). 
JM thanks for support by FAPESP (2012/24392-2). 
MA has been supported by the Australian Research Council (grants FL110100012 and DP120100991).
MB is supported by the National Science Foundation (NSF) Graduate Research Fellowships Program (grant no. DGE-1144082).
JB and MB acknowledge support for this work from the NSF (grant no. AST-1313119). 
JB is also supported by the Alfred P. Sloan Foundation and the David and Lucile Packard Foundation.
\end{acknowledgements}


\Online

\begin{appendix}
\newgeometry{left=2.5cm,bottom=2cm}
\small
\setcounter{table}{1}
\begin{center}
\begin{longtable}{lcccccccccccc}
\caption{[Y/H], [Mg/H], stellar parameters and age for the 88 solar twins sample. The binary stars are identified by {\it *}.}\\
\hline \hline \\[-2ex]
\multicolumn{1}{c}{Star} &
\multicolumn{1}{c}{[Y/H]} &
\multicolumn{1}{c}{error} &
\multicolumn{1}{c}{[Mg/H]} &
\multicolumn{1}{c}{error} &
\multicolumn{1}{c}{[Fe/H]} &
\multicolumn{1}{c}{error} &
\multicolumn{1}{c}{T$\rm_{eff}$} &
\multicolumn{1}{c}{error} &
\multicolumn{1}{c}{log $g$} &
\multicolumn{1}{c}{error} &
\multicolumn{1}{c}{Age} &
\multicolumn{1}{c}{error}\\
 
\hline
\endfirsthead

\multicolumn{13}{c}{\footnotesize{{\slshape{{\tablename} \thetable{}}} - Continuation}}\\

\hline \hline\\[-2ex]
 
\multicolumn{1}{c}{{Star}} &
\multicolumn{1}{c}{{[Y/H]}} &
\multicolumn{1}{c}{{error}} &
\multicolumn{1}{c}{{[Mg/H]}} &
\multicolumn{1}{c}{{error}} &
\multicolumn{1}{c}{{[Fe/H]}} &
\multicolumn{1}{c}{{error}} &
\multicolumn{1}{c}{T$\rm_{eff}$} &
\multicolumn{1}{c}{error} &
\multicolumn{1}{c}{log {\it g}} &
\multicolumn{1}{c}{error} &
\multicolumn{1}{c}{{[Age]}} &
\multicolumn{1}{c}{{error}} \\ 

\hline 
\endhead

\hline \multicolumn{9}{l}{{Continued on next page}} \\ 
\endfoot
\hline 

\endlastfoot

HIP 10175  & 0.043 & 0.010 & -0.048 & 0.005 &  -0.007  & 0.005  & 5738&7&4.51&0.01    & 1.815 & 0.652\\ 
HIP 101905 & 0.129 & 0.009 & 0.031 & 0.005 &  0.057  &  0.006  & 5890&6&4.47&0.02    & 1.589 & 0.685\\
HIP 102040 & -0.033 & 0.006 & -0.100 & 0.013 &  -0.093  & 0.006  & 5838&6&4.48&0.02    & 2.423 & 0.912\\
HIP 102152 & -0.083 & 0.020 & 0.000 & 0.018 &  -0.020  & 0.005  & 5718&5&4.40&0.02    & 6.918 & 0.689\\
HIP 10303  & 0.112 & 0.016 & 0.123 & 0.022 &  0.106  &  0.004  & 5725&4&4.40&0.01    & 5.477 & 0.561\\
HIP 103983* & 0.031 & 0.014 & -0.030 & 0.020 & -0.048  & 0.008  & 5752&10&4.51&0.02   & 2.077 & 0.859\\
HIP 104045 & 0.092 & 0.010 & 0.031 & 0.006 &  0.045  &  0.005  & 5831&6&4.47&0.02    & 2.293 & 0.833\\
HIP 105184 & 0.109 & 0.013 & -0.043 & 0.008 &  -0.002  & 0.009  & 5833&11&4.504&0.02  & 0.604 & 0.445\\
HIP 108158 & 0.008 & 0.017 & 0.200 & 0.031 &  0.067  &  0.008  & 5687&7&4.34&0.02    & 8.364 & 0.477\\
HIP 108468 & -0.233 & 0.011 & -0.071 & 0.026 &  -0.111  & 0.006  & 5829&7&4.33&0.02    & 7.562 & 0.397\\
HIP 108996 & 0.141 & 0.009 & -0.013 & 0.014 &  0.064  &  0.013  & 5847&17&4.503&0.03  & 0.978 & 0.700\\
HIP 109110* & 0.194 & 0.018 & 0.002 & 0.019 &  0.035  &  0.014  & 5787&17&4.50&0.04  & 2.335 & 1.212\\
HIP 109821 & -0.181 & 0.010 & -0.033 & 0.011 &  -0.115  & 0.005  & 5746&7&4.31&0.02    & 9.301 & 0.390\\
HIP 114615 & -0.007 & 0.011 & -0.107 & 0.007 &  -0.077  & 0.008  & 5816&9&4.52&0.02    & 1.050 & 0.710\\
HIP 115577 & -0.066 & 0.020 & 0.153 & 0.012 &  0.036  &  0.008  & 5699&9&4.25&0.03    & 9.501 & 0.342\\
HIP 116906 & -0.055 & 0.017 & 0.056 & 0.009 &  0.010  &  0.005  & 5792&6&4.37&0.02    & 6.463 & 0.441\\
HIP 117367 & -0.018 & 0.003 & 0.038 & 0.009 &  0.044  &  0.007  & 5871 & 8 & 4.32 & 0.02    & 5.942 & 0.395\\
HIP 118115 & -0.153 & 0.007 & -0.002 & 0.011 &  -0.017  & 0.006  & 5808&7&4.28 & 0.02    & 7.791 & 0.324\\
HIP 11915  & -0.032 & 0.006 & -0.074 & 0.011 &  -0.059  & 0.004  & 5760&4&4.46 & 0.01    & 4.157 & 0.647\\
HIP 14501  & -0.219 & 0.010 & 0.016 & 0.015 &  -0.133  & 0.005  & 5728&7&4.29 & 0.02 & 9.926 & 0.374\\
HIP 14614  & -0.093 & 0.009 & -0.117 & 0.010  & -0.099   &0.008  & 5784&9&4.42 & 0.03& 5.823 & 1.016\\
HIP 14623  & 0.144 & 0.013 & 0.061 & 0.013 &  0.106  &  0.01   & 5769&13&4.52&0.02   & 1.137 & 0.642\\
HIP 15527  & -0.203 & 0.010 & -0.036 & 0.005 &  -0.051  & 0.005  & 5785&5&4.32&0.01    & 7.924 & 0.320\\
HIP 18844  & -0.072 & 0.007 & 0.068 & 0.005 &  0.016  &  0.004  & 5736&5&4.36&0.02    & 7.456 & 0.427\\
HIP 1954   & -0.049 & 0.010 & -0.089 & 0.009 &  -0.068  & 0.006  & 5717&5&4.46&0.02    & 4.872 & 0.965\\
HIP 19911*  &0.049 & 0.035 & -0.100 & 0.023 & -0.070   & 0.011 & 5764&12&4.47&0.04   & 4.004 & 1.466\\
HIP 21079  & -0.008 & 0.016 & -0.138 & 0.014 &  -0.070  &  0.008 & 5846&11&4.50&0.03   & 1.663 & 0.977\\
HIP 22263  & 0.112 & 0.014 & -0.026 & 0.019 &  0.030  &   0.007 & 5840&8&4.50&0.02    & 1.074 & 0.762\\
HIP 22395** & 0.054 & 0.012 & 0.068 & 0.020 &  0.084  &   0.008 & 5789&8&4.43&0.02    & 3.934 & 0.853\\
HIP 25670  & 0.095 & 0.010 & 0.040 & 0.012 &  0.057  &   0.005 & 5771&5&4.44&0.02    & 4.120 & 0.768\\
HIP 28066  & -0.227 & 0.001 & 0.033 & 0.006 &  -0.128  &  0.004 & 5733&5&4.29&0.01    & 9.859 & 0.295\\
HIP 29432  & -0.120 & 0.007 & -0.111 & 0.017 &  -0.096  &  0.005 & 5758&5&4.44&0.01    & 5.508 & 0.710\\
HIP 29525  & 0.039 & 0.012 & -0.078 & 0.019  & -0.022   & 0.007 & 5737&7&4.49&0.02    & 2.827 & 1.056\\
HIP 30037*  &-0.010 & 0.015 & 0.015 & 0.015 & -0.011   & 0.004 & 5668&5&4.42&0.01    & 6.960 & 0.624\\
HIP 30158  & -0.003 & 0.021 & 0.041 & 0.007 &  0.003  &   0.006 & 5702&5&4.46&0.02    & 4.570 & 0.981\\
HIP 30344  & 0.122 & 0.006 & -0.002 & 0.021 &  0.063  &   0.007 & 5750&9&4.50&0.02    & 1.924 & 0.826\\
HIP 30476  & -0.138 & 0.005 & 0.077 & 0.010  & -0.022   & 0.004 & 5710&5&4.26&0.01    & 9.689 & 0.273\\
HIP 30502  & -0.081 & 0.014 & -0.051 & 0.009 &  -0.076  &  0.006 & 5721&6&4.41&0.02    & 7.007 & 0.679\\
HIP 3203   & 0.000 & 0.015 & -0.148 & 0.014 &  -0.087  &  0.008 & 5850&10&4.52&0.02   & 0.987 & 0.662\\
HIP 33094  & -0.032 & 0.010 & 0.182 & 0.016 &  0.043  &   0.005 & 5662&7&4.16&0.02    &10.092 & 0.265\\
HIP 34511  & -0.052 & 0.003 & -0.102 & 0.008 &  -0.103  &  0.006 & 5819&6&4.47&0.02    & 3.373 & 0.889\\
HIP 36512  & -0.148 & 0.005 & -0.100 & 0.010 &  -0.117  &  0.004 & 5737&4&4.41&0.01    & 7.185 & 0.500\\
HIP 36515  & 0.044 & 0.009 & -0.100 & 0.019  & -0.021   & 0.009 & 5847&12&4.54&0.02   & 0.633 & 0.464\\
HIP 38072  & 0.088 & 0.018 & 0.035 & 0.009 &  0.058  &   0.007 & 5849&8&4.49&0.02    & 1.306 & 0.724\\
HIP 40133  & 0.088 & 0.018 & 0.159 & 0.012 &  0.128  &   0.004 & 5755&4&4.37&0.01    & 5.500 & 0.389\\
HIP 41317  & -0.161 & 0.004 & -0.038 & 0.005 &  -0.068  &  0.004 & 5700&5&4.38&0.01    & 8.224 & 0.468\\
HIP 42333  & 0.210 & 0.006 & 0.055 & 0.007 &  0.138  &   0.008 & 5848&8&4.50&0.02    & 1.011 & 0.518\\
HIP 43297*  &0.151 & 0.011 & 0.037 & 0.014 & 0.083   &  0.006 & 5702&5&4.46&0.01    & 3.840 & 0.738\\
HIP 44713  & -0.027 & 0.010 & 0.095 & 0.009 &  0.088  &   0.005 & 5768&6&4.28&0.01    & 7.581 & 0.288\\
HIP 44935  & 0.001 & 0.013 & 0.066 & 0.012 &  0.058  &   0.005 & 5782&5&4.37&0.01    & 6.215 & 0.434\\
HIP 44997  & 0.029 & 0.016 & -0.019 & 0.025 &  -0.023  &  0.005 & 5731&5&4.47&0.02    & 3.876 & 0.919\\
HIP 4909   & 0.108 & 0.021 & -0.065 & 0.011  & 0.028   &  0.008 & 5854&10&4.50&0.02   & 1.232 & 0.770\\
HIP 49756  & 0.056 & 0.008 & 0.014 & 0.012 &  0.043  &   0.004 & 5795&4&4.42&0.01    & 4.618 & 0.573\\
HIP 5301   & -0.093 & 0.009 & -0.082 & 0.003 &  -0.064  &  0.004 & 5728&5&4.42&0.02    & 6.488 & 0.670\\
HIP 54102*  &0.089 & 0.007 & -0.062 & 0.008 & -0.014   & 0.007 & 5820&9&4.51&0.02    & 1.107 & 0.698\\
HIP 54287  & 0.069 & 0.015 & 0.148 & 0.013 &  0.118  &   0.004 & 5727&4&4.36&0.01    & 6.340 & 0.398\\
HIP 54582*  &-0.210 & 0.011 & -0.069 & 0.011 & -0.080   & 0.005 & 5875&7&4.27&0.02    & 7.276 & 0.312\\
HIP 55409  & -0.078 & 0.009 & -0.070 & 0.014 &  -0.080  &  0.006 & 5700&6&4.40&0.02    & 7.655 & 0.650\\
HIP 62039*  &0.041 & 0.016 & 0.120 & 0.010 & 0.088   &  0.005 & 5753&6&4.35&0.02    & 6.725 & 0.441\\
HIP 6407*   &-0.005 & 0.011 & -0.114 & 0.009 & -0.068   & 0.007 & 5764&8&4.52&0.01    & 1.488 & 0.656\\
HIP 64150*  &0.339 & 0.011 & 0.087 & 0.015 & 0.030   &  0.007 & 5747&6&4.39&0.02    & 6.406 & 0.656\\
HIP 64673*  &-0.091 & 0.009 & -0.011 & 0.010 & -0.030   & 0.007 & 5918&8&4.35&0.02    & 5.224 & 0.554\\
HIP 64713  & -0.025 & 0.014 & -0.068 & 0.015 &  -0.067  &  0.007 & 5767&8&4.46&0.02    & 4.261 & 1.096\\
HIP 65708  & -0.132 & 0.005 & 0.039 & 0.016 &  -0.066  &  0.006 & 5755&6&4.25&0.02    & 9.410 & 0.284\\
HIP 67620*  &0.028 & 0.017 & -0.003 & 0.016 & -0.018   & 0.009 & 5670&9&4.41&0.03    & 7.176 & 1.077\\
HIP 68468  & 0.016 & 0.005 & 0.105 & 0.016 &  0.054  &   0.005 & 5845&6&4.37&0.02    & 5.334 & 0.467\\
HIP 69645  & -0.034 & 0.020 & -0.043 & 0.010  & -0.045   & 0.006 & 5743&6&4.44&0.02    & 5.273 & 0.852\\
HIP 72043*  &-0.153 & 0.014 & 0.023 & 0.015 & -0.034   & 0.007 & 5842&8&4.35&0.02    & 6.419 & 0.468\\
HIP 73241*  &0.015 & 0.036 & 0.185 & 0.011 & 0.082   &  0.007 & 5669&8&4.27&0.02    & 9.384 & 0.346\\
HIP 73815  & -0.058 & 0.018 & 0.037 & 0.011 &  0.004  &   0.005 & 5788&6&4.37&0.02    & 6.566 & 0.462\\
HIP 74389  & 0.127 & 0.003 & 0.059 & 0.015 &  0.077  &   0.004 & 5844&5&4.49&0.01    & 1.005 & 0.484\\
HIP 74432** & 0.011 & 0.009 & 0.168 & 0.011 &  0.037  & 0.007 & 5684&8&4.25&0.02    & 9.768 & 0.312\\
HIP 7585   & 0.127 & 0.009 & 0.087 & 0.005 &  0.095  &   0.005 & 5831&5&4.43&0.01    & 3.291 & 0.508\\
HIP 76114  & -0.035 & 0.010 & -0.030 & 0.011 &  -0.037  &  0.006 & 5733&6&4.42&0.02    & 6.151 & 0.816\\
HIP 77052** & 0.141 & 0.016 & 0.052 & 0.013 &  0.036  & 0.006 & 5683&5&4.48&0.02    & 3.665 & 0.906\\
HIP 77883  & -0.079 & 0.024 & 0.018 & 0.008 &  -0.006  &  0.006 & 5690&6&4.4&0.02     & 7.240 & 0.678\\
HIP 79578*  & 0.071 & 0.008 & 0.043 & 0.012 & 0.057   &  0.005 & 5820&5&4.47&0.01    & 2.170 & 0.778\\
HIP 79672  & 0.090 & 0.010 & 0.054 & 0.019 &  0.056  &   0.003 & 5814&3&4.45&0.01    & 3.090 & 0.391\\
HIP 79715  & -0.125 & 0.004 & -0.025 & 0.010  & -0.041   & 0.005 & 5803&6&4.38&0.02    & 6.471 & 0.462\\
HIP 81746* & -0.167 & 0.012 & -0.037 & 0.016 & -0.086   & 0.004 & 5715&5&4.40&0.02    & 7.526 & 0.582\\
HIP 83276** & -0.242 & 0.004 & -0.079 & 0.008 &  -0.089 & 0.006 & 5885&8&4.22&0.02    & 7.543 & 0.267\\
HIP 85042  & -0.001 & 0.013 & 0.034 & 0.012 &  0.015  &   0.004 & 5694&5&4.41&0.02    & 6.662 & 0.617\\
HIP 8507   & -0.046 & 0.006 & -0.114 & 0.012 &  -0.096  &  0.006 & 5725&6&4.49&0.02    & 3.625 & 0.943\\
HIP 87769*  & 0.063 & 0.010 & 0.005 & 0.006 & 0.000   &  0.006 & 5807&6&4.40&0.02    & 5.145 & 0.687\\
HIP 89650  & -0.031 & 0.007 & -0.057 & 0.014 &  0.000  &   0.005 & 5841&5&4.44&0.02    & 3.824 & 0.755\\
HIP 9349   & 0.050 & 0.015 & -0.042 & 0.015 &  0.009  &   0.007 & 5810&8&4.50&0.02    & 1.429 & 0.758\\
HIP 95962** & -0.001 & 0.010 & 0.047 & 0.012 &  0.023  &   0.005 & 5806&5&4.44&0.02    & 3.820 & 0.776\\
HIP 96160  & 0.020 & 0.004 & -0.059 & 0.012 &  -0.053  &  0.007 & 5781&8&4.50&0.02    & 2.165 & 0.779\\
                    
\label{bin}                    
\end{longtable}
\end{center}
* Spectroscopic binary star; ** Visual binary star
\restoregeometry
\end{appendix}
\end{document}